\documentclass[11pt,oneside]{article}
\pdfoutput=1
\usepackage{setspace,graphicx,amssymb,amsmath,latexsym,amsfonts,amscd,amsthm,multirow,ctable,mathdots,caption,array,diagbox,mathtools}
\usepackage{tabularx,cite,mathrsfs}
\usepackage{authblk}
\usepackage{color}

\usepackage[margin=2cm]{caption}

\usepackage{fullpage}
\usepackage{stmaryrd}
\usepackage{rotating}

\usepackage{hyperref}

\newcommand{\bea}{\begin{eqnarray}}
\newcommand{\eea}{\end{eqnarray}}
\newcommand{\bee}{\begin{eqnarray*}}
\newcommand{\eee}{\end{eqnarray*}}
\newcommand{\al}{\begin{align*}}
\newcommand{\eal}{\end{align*}}
\newcommand{\be}{\begin{equation}}
\newcommand{\ee}{\end{equation}}
\newcommand{\bem}{\begin{pmatrix}}
\newcommand{\eem}{\end{pmatrix}}

\newcommand{\mc}{\mathcal}
\newcommand{\bb}{\mathbb}

\def\th{\theta}

\def\Tr{\text{Tr}}

\def\el{\in}
\def\({\left(}
\def\){\right)}
\def\nn{\nonumber}
\numberwithin{equation}{section}

\begin{document}

\setstretch{1.4}

\title{\vspace{-65pt}
\vspace{20pt}
    \textsc{%\huge{
    A Refined Count of BPS States in the D1/D5 System     }
}

\author[]{Nathan Benjamin\thanks{nathansbenjamin@gmail.com}
}

\affil[]{Stanford Institute for Theoretical Physics, Department of Physics\\
Stanford University, Stanford, CA 94305, USA}

\date{}

\maketitle

\vspace{-1em}

\abstract{

We examine the low-lying quarter BPS spectrum of a 2d conformal field theory with target Sym$^N(K3)$ at various points in the moduli space, and look at a more refined count than the ordinary elliptic genus. We compute growth of the spectrum at both the symmetric orbifold point, as well as at the supergravity point in the moduli space. Finally we do a decomposition of the spectra into $\mc{N}=4$ characters and discuss possible relations to interesting symmetry groups. A similar analysis is done with $T^4$.
}

\clearpage

\tableofcontents

\section{Introduction}
\label{sec:intro}

One of the earliest examples of the AdS/CFT correspondence \cite{Maldacena:1997re} is the D1/D5 system on $K3$ or $T^4$. There, we expect that a marginal deformation of a 2d CFT with target Sym$^N(K3)$ is dual to large radius gravity on AdS$_3 \times S^3 \times K3$ (and similarly for $T^4$). In particular, there should be a region in the conformal moduli space where the low-lying CFT spectrum matches the 6d supergravity Kaluza-Klein spectrum. A very nontrivial check of this conjecture was done in \cite{deBoer:1998kjm, deBoer:1998us} where the elliptic genus of Sym$^N(K3)$ was calculated at the symmetric orbifold point, and compared to the supergravity elliptic genus, and the two were found to match. 

More generally, one can look at the growth of the low-lying states in a 2d CFT and see if it is consistent for it to be dual to large-radius gravity. In particular, elliptic genus of a purported 2d CFT dual to large-radius gravity must have the number of low-lying states grow subexponentially with dimension \cite{Benjamin:2015hsa}. The growth of the low-lying states in the elliptic genus of Sym$^N(K3)$ is given by \cite{Benjamin:2015vkc}
\be
\rho^{\text{EG}}_{\text{Sym}^N(K3)}(n) \sim N \text{exp}\(\sqrt{48} \pi \sqrt{n}\)
\label{eq:pig}
\ee
where $\rho^{\text{EG}}_{\text{Sym}^N(K3)}(n)$ counts states $n$ above the vacuum in the NS-sector elliptic genus of Sym$^N(K3)$, and crucially $1 \ll n \ll c=6N$.\footnote{The divergence as $N$ goes to infinity is simply due to the large number of right-moving Ramond ground states that all contribute to the genus.}

In this paper, instead of simply computing the elliptic genus (a signed count of quarter-BPS states) of Sym$^N(K3)$, we will compute the recently introduced ``Hodge elliptic genus" \cite{Kachru:2016igs}, which counts quarter-BPS states without cancellations. We compute this object at various points in the symmetric orbifold moduli space, and see the growth of the low-lying states. In particular, we will find that at the supergravity point, the Hodge elliptic genus grows as
\be
\rho^{\text{HEG}}_{\text{sugra},K3}(n) \sim N \exp{\(\frac{4 \sqrt{2} \pi}{3^{3/4}}n^{3/4}\)}
\label{eq:34}
\ee
which is parametrically faster than (\ref{eq:pig}). Thus we expect a very large number of cancellations in the quarter-BPS spectrum, even at the supergravity point in moduli space. From dimensional analysis arguments, we expect the total partition function of the six-dimensional supergravity to grow as
\be
\rho^{\text{PF}}_{\text{sugra},K3}(n) \sim \text{exp}\(n^{5/6}\).
\label{eq:56}
\ee
The mismatch between (\ref{eq:56}) and (\ref{eq:34}) can be understood as restricting our attention to BPS states; the mismatch between (\ref{eq:34}) and (\ref{eq:pig}) can be understood as cancellations between said BPS states. We pause to emphasize that the refined count of states we propose is counting not black hole states in the gravity description, but low-lying KK modes. Indeed, in the known examples where the elliptic genus already gets the black hole entropy correct (such as the D1/D5 system on $K3$ \cite{Strominger:1996sh}), the Hodge elliptic genus and the elliptic genus must have the same parametric growth when counting black hole states.

The organization of this note is as follows. In Section \ref{sec:dog} we will study the growth of the Hodge elliptic genus for both Sym$^N(K3)$ and Sym$^N(T^4)$ at the symmetric orbifold point and at the supergravity point. In Section \ref{sec:cat} we decompose the various Hodge elliptic genera calculated into $\mc{N}=4$ characters and see how the decomposition changes at different points in the moduli space. Suggestive relations to symmetry groups including the sporadic finite group $M_{22}$ are found in the decomposition. In Section \ref{sec:fish} we give some concluding remarks. 

\section{Growth of BPS States}
\label{sec:dog}

The Hodge elliptic genus, introduced in \cite{Kachru:2016igs}, is defined for theories with at least $\mc{N}=(2,2)$ supersymmetry as
\be
Z_{HEG}(\tau, z, \nu) = \Tr_{\text{RR}}\((-1)^{F_L+F_R}q^{L_0-\frac{c}{24}}y^{J_0} u^{\bar{J_0}}\)\bigg|_{\bar h = \frac{c}{24}}
\label{eq:HEG}
\ee
where trace is taken only over right-moving Ramond ground states, which have $\bar{h}=\frac{c}{24}$. Here and throughout this paper, we define $q = e^{2\pi i \tau}, y = e^{2\pi i z}, u = e^{2 \pi i \nu}$. By definition, this object only counts quarter-BPS states, since we only take contributions from right-moving supersymmetric states. Note that the Hodge elliptic genus is not an index; as we vary around points in the CFT moduli space, various short multiplets can combine to form long multiplets in a way that does not leave (\ref{eq:HEG}) invariant. If we take $y=u=-1$ in (\ref{eq:HEG}), then this reduces to
\be
Z_{HEG}\(\tau, 1/2, 1/2\) = \Tr_{\text{RR}}\(q^{L_0-\frac{c}{24}}\)\big|_{\bar h = \frac{c}{24}},
\label{eq:horses}
\ee
in other words a simple count of the quarter-BPS spectrum of the theory. 

\subsection{$K3$}
\label{sec:k3}

We will first investigate the growth of coefficients of the Hodge elliptic genus in Sym$^N(K3)$. We study the growth at three points in the moduli space: at the symmetric orbifold point of the $T^4/\bb Z_2$ orbifold; at the symmetric orbifold point of a generic $K3$ surface; and at the supergravity description. Much of this analysis will be quite similar to that in \cite{Benjamin:2015vkc}.

\subsubsection{Sym$^N(T^4/\bb Z_2)$}
\label{sec:t4z2}

We will first consider a $T^4/\bb Z_2$ orbifold as our $K3$ surface. The Hodge elliptic genus of this is easily calculated from free field theory; it is calculated in \cite{Kachru:2016igs} and is given by
\begin{align}
Z_{\text{HEG}, T^4/\bb{Z}_2}(\tau,z,\nu) &= 8\(\(\frac{\th_1(\tau,z)}{\th_1(\tau)} u_-\)^2 + \(\frac{\th_2(\tau,z)}{\th_2(\tau)} u_+\)^2 + \(\frac{\th_3(\tau,z)}{\th_3(\tau)} \)^2 + \(\frac{\th_4(\tau,z)}{\th_4(\tau)} \)^2\) \nn \\
&= u^{-1} y^{-1} + u y^{-1} + u^{-1} y + u y + 20 + \big(4 u^{-1} y^{-1} + 4 u y^{-1} + 4 u^{-1} y + 4 u y \nn \\ &\phantom{aaaa} - 136 y - 136y^{-1} + 20 y^2 + 20y^{-2} - 8u - 8u^{-1} + 232\big)q + \mc{O}(q^2) \nn\\
&:= \sum_{m, \ell, \ell'} c_{T^4/\bb{Z}_2}(m, \ell, \ell') q^m y^\ell u^{\ell'}
\end{align}
where $u_{\pm} := \frac12\(u^{1/2} \pm u^{-1/2}\)$ and the theta functions are defined in Appendix \ref{app:A}.

We can use a modification of the DMVV formula \cite{Dijkgraaf:1996xw, Kachru:2016igs} to get the Hodge elliptic genus growth at the symmetric orbifold point. In particular, we have
\be
\sum_{n\geq0} Z_{\text{HEG}, \text{Sym}^n(T^4/\bb{Z}_2)}(\tau, z, \nu) p^n = \prod_{\substack{n>0 \\ m\geq0 \\ \ell, \ell'}}\frac{1}{(1-p^nq^my^\ell u^{\ell'})^{c_{T^4/\bb{Z}_2}(nm,\ell,\ell')}}.
\label{eq:dmvv}
\ee
We can spectral flow to the NS-R sector\footnote{For convenience, we use a convention where the NS vacuum is at $q^0$, not $q^{-c/24}=q^{-n}$.} to get
\begin{align}
\sum_{n\geq0} Z_{\text{HEG}, \text{Sym}^n(T^4/\bb{Z}_2)}^{\text{NS}}(\tau, z, \nu) p^n &= \prod_{\substack{n>0 \\ m\geq0 \\ \ell, \ell'}}\frac{1}{(1-p^nq^{m+\ell/2+n/2}y^{\ell+n} u^{\ell'})^{c_{T^4/\bb{Z}_2}(nm,\ell,\ell')}} \nn\\
&= \prod_{\substack{n>0 \\ m\el \frac{\bb Z}2 \\ m-\frac{\ell}2\el\bb{Z}}} \frac{1}{(1-p^nq^my^\ell u^{\ell'})^{c_{T^4/\bb{Z}_2}(nm-\frac{n\ell}2,\ell-n,\ell')}}.
\label{eq:dmvvns}
\end{align}
In order to reproduce the growth of quarter-BPS states with no cancellations, we can simply set $y=u=-1$, as in (\ref{eq:horses}). After doing so, note that the only term in the product we get with no $q$ dependence occurs at $n=1, m=0, \ell=0, \ell'=\pm1$ in the product. Thus we get
\be
\sum_{n\geq0} Z_{\text{HEG}, \text{Sym}^n(T^4/\bb{Z}_2)}^{\text{NS}}(\tau, 1/2, 1/2) p^n = \frac{1}{(1+p)^2}\prod_{\substack{n>0 \\ m\el \frac{\bb Z}2, m>0 \\ m-\frac{\ell}2\el\bb{Z}}} \frac{1}{(1-p^nq^m(-1)^{\ell+\ell'} )^{c_{T^4/\bb{Z}_2}(nm-\frac{n\ell}2,\ell-n,\ell')}}.
\label{eq:lion}
\ee
The divergence of the BPS spectrum as $n\rightarrow\infty$ due to right-moving Ramond ground states can be seen in the 
\be
\frac{1}{(1+p)^2} = 1 - 2p + 3p^2 - 4p^3 + \ldots
\ee
prefactor of (\ref{eq:lion}). At large $N$, the growth of states of Sym$^N(T^4/\bb{Z}_2)$ will therefore scale linearly with $N$; to extract the coefficient, we can set $p=-1$ in the product in (\ref{eq:lion}). In other words, at large $N$ we have
\begin{align}
\frac1NZ_{\text{HEG}, \text{Sym}^N(T^4/\bb{Z}_2)}^{\text{NS}}(\tau, 1/2, 1/2) &= \prod_{\substack{n>0 \\ m\el \frac{\bb Z}2, m>0 \\ m-\frac{\ell}2\el\bb{Z}}} \frac{1}{(1-q^m(-1)^{n+\ell+\ell'} )^{c_{T^4/\bb{Z}_2}(nm-\frac{n\ell}2,\ell- n,\ell')}} + \mc{O}(q^{N/4}) \nn \\
&= \prod_{m\el \frac{\bb Z}2, m>0} \frac{(1+q^m)^{f_{T^4/\bb Z_2}(m)}}{(1-q^m)^{g_{T^4/\bb Z_2}(m)}} + \mc{O}(q^{N/4})
\label{eq:tiger}
\end{align}
where $f_{T^4/\bb Z_2}(m)$ and $g_{T^4/\bb Z_2}(m)$ are defined as
\begin{align}
f_{T^4/Z_2}(m) &= \sum_{\substack{n>0, n\el \bb{Z} \\ \ell \equiv 2m \text{~(mod~}2) \\ \ell' \equiv 2m+n+1 \text{~(mod~}2)}} -c_{T^4/\bb{Z}_2}(nm-\frac{n\ell}2,\ell- n,\ell')\nn \\
g_{T^4/Z_2}(m) &= \sum_{\substack{n>0, n\el \bb{Z} \\ \ell \equiv 2m \text{~(mod~}2) \\ \ell' \equiv 2m+n \text{~(mod~}2)}} c_{T^4/\bb{Z}_2}(nm-\frac{n\ell}2,\ell- n,\ell').
\label{eq:pony}
\end{align}
Note if we consider $g_{T^4/\bb Z_2}(m)-f_{T^4/\bb Z_2}(m)$, we are summing over $\ell'$, effectively setting $u=1$ in the Hodge elliptic genus, which reproduces the $K3$ elliptic genus. Thus
\be
g_{T^4/\bb Z_2}(m) - f_{T^4/\bb Z_2}(m) = \sum_{\substack{n>0 , n \el \bb{Z} \\ \ell \equiv 2m \text{~(mod~}2)}} c_{K3}^{\text{EG}}(nm-\frac{n\ell}2,\ell-n)
\label{eq:us}
\ee
where $c_{K3}^{\text{EG}}$ measures the coefficients of the elliptic genus of $K3$, not the Hodge elliptic genus. In \cite{Benjamin:2015vkc}, the RHS of (\ref{eq:us}) was calculated to be 44 for all half-integer $m$, and 28 for integer $m$. The asymptotic growth of $f_{T^4/\bb Z_2}(m)$ and $g_{T^4/\bb Z_2}(m)$ then are the same. Moreover, it can be shown that if $f_{T^4/\bb Z_2}(m) + g_{T^4/\bb Z_2}(m)$ grow as $m^p$ for some $p$, then the $q^n$ coefficient of (\ref{eq:tiger}) grows as $\text{exp}\(n^{\frac{p+1}{p+2}}\)$; but if it grows faster than polynomial in $m$, then (\ref{eq:tiger}) exhibits Hagedorn growth\footnote{We know it cannot grow superexponentially since the partition function itself at any symmetric orbifold point has a Hagedorn density of states \cite{Keller:2011xi, Haehl:2014yla, Belin:2014fna, Belin:2015hwa}, which provides an upper bound.} (see Appendix \ref{app:B}).

We can of course do a very similar analysis for any symmetric orbifold Sym$^N(X)$. We would then define the analogous functions $f_X(m)$ and $g_X(m)$ as in (\ref{eq:pony}):
\begin{align}
f_{X}(m) &= \sum_{\substack{n>0, n\el \bb{Z} \\ \ell \equiv 2m \text{~(mod~}2) \\ \ell' \equiv 2m+n+1 \text{~(mod~}2)}} -c_{X}(nm-\frac{n\ell}2,\ell- \frac{c n}6,\ell')\nn \\
g_{X}(m) &= \sum_{\substack{n>0, n\el \bb{Z} \\ \ell \equiv 2m \text{~(mod~}2) \\ \ell' \equiv 2m+n \text{~(mod~}2)}} c_{X}(nm-\frac{n\ell}2,\ell-\frac{c n}6,\ell').
\label{eq:ponyup}
\end{align} 
where $c_X$ are the coefficients of the seed theory $X$ and $c$ is the central charge of the seed theory $X$. At large $N$, the number of BPS states $n$ above the vacuum is similarly given by the $q^{n}$ term of 
\be
\prod_{m\el \frac{\bb Z}2, m>0} \frac{(1+q^m)^{f_{X}(m)}}{(1-q^m)^{g_{X}(m)}}.
\label{eq:fishfish}
\ee

In fact, we can show that at any symmetric orbifold point, $f_X(m)$ will grow as $e^{2\pi m}$, which will give us a Hagedorn density of BPS states. In particular, every term in both sums of (\ref{eq:ponyup}) is manifestly nonnegative. First, $f_X(m)$ only gets contributions from terms with odd eigenvalue under $J_0 + \bar{J_0}$, which therefore gets a minus sign under the $(-1)^F$ in (\ref{eq:HEG}) that cancels the sign in the definition of $f_X(m)$. Likewise $g_X(m)$ only gets contributions from terms with even eigenvalue under the same operator, so each term comes positive. Thus we can get no cancellations, and can put a lower bound on the growth by looking at one term in the sum. In \cite{Benjamin:2015vkc}, this was estimated from the Cardy formula of the seed theory, which gives $f_X(m)$ growing as $e^{2\pi m}$ at large $m$. Note that unlike in \cite{Benjamin:2015vkc}, no cancellations are allowed in $f_X(m)$ and $g_X(m)$; thus, we cannot get sub-Hagedorn growth. %check this e^{\sqrt{m}} claim since it's different from 1512.00010

\begin{figure}[!h]
\centering
\includegraphics[width=0.85\textwidth]{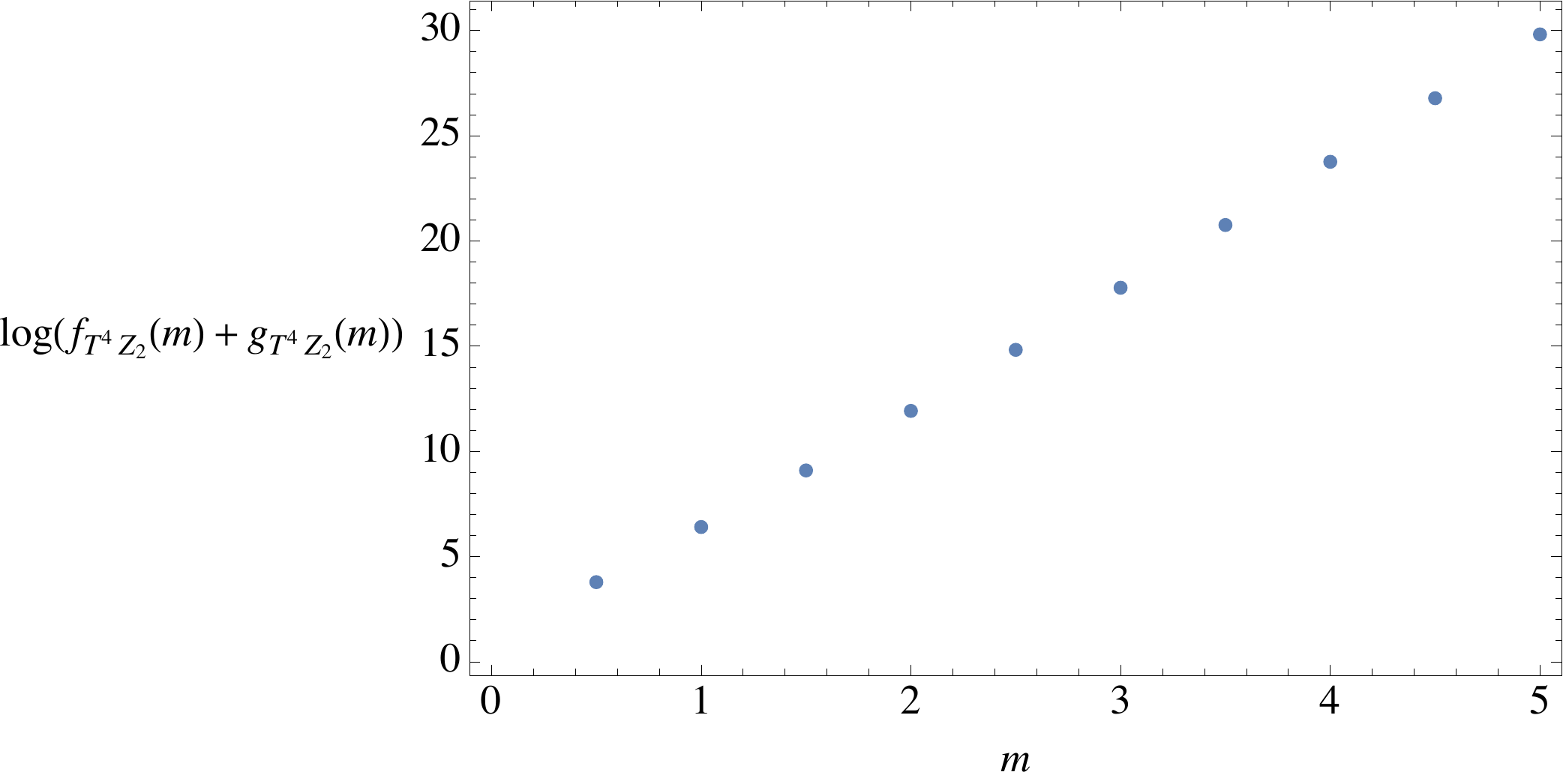}
\caption{A plot of $\log{(f_{T^4/\bb Z_2}(m)+g_{T^4/\bb Z_2}(m))}$ up to $m=5$ for $T^4/\bb{Z}_2$. Note the exponential growth in $m$.}
\label{fig:toad}
\end{figure}

To illustrate, we show the first few values of $f_{T^4/\bb Z_2}(m)$ and $g_{T^4/\bb Z_2}(m)$ in Table \ref{tab:t4z2}, and plot the sum in Figure \ref{fig:toad}.
\begin{table}[h]
\begin{center}
\begin{tabular}{| c || c  c  c  c  c  c  c  c |}
\hline
$m$ & $\frac12$ & 1 & $\frac32$ & 2 & $\frac52$ & 3 & $\frac72$ & 4 \\ \hline %& $\frac92$ & 5 \\ \hline
$f_{T^4/\bb Z_2}(m)$ & 0 & 288 & 4416 & 75168 & 1370688 & 26195808 & 516627840 & 10420480416 \\  %& 213762288384 & 4443111232608 \\ \hline
$g_{T^4/\bb Z_2}(m)$ & 44 & 316 & 4460 & 75196 & 1370732 & 26195836 & 516627884 & 10420480444 \\  \hline %& 213762288428 & 4443111232636 \\
\end{tabular}
\end{center}
\caption{First few values of $f_{T^4/\bb Z_2}(m)$ and $g_{T^4/\bb Z_2}(m)$, defined in (\ref{eq:pony}).}
\label{tab:t4z2}
\end{table}

\subsubsection{Sym$^N(\text{Generic~}K3)$}

The $T^4/\bb{Z}_2$ theory is at a very special point in the $K3$ moduli space. We can also consider the symmetric orbifold of a generic $K3$ surface. The Hodge elliptic genus for a generic $K3$ was recently computed in \cite{Wendland}, by making the simple assumption that at a generic point in the $K3$ moduli space, the chiral algebra should not enhance: for all integers $n>0$, there should be no states with $h=n, \bar{h}=0$ (or the reverse). 

For $K3$, it turns out that this assumption, combined with knowledge of the elliptic genus, is enough to fix the Hodge elliptic genus. It is given by\cite{Wendland}
\begin{align}
Z_{\text{HEG}, K3}(\tau, z, \nu) &= (2-u-u^{-1})\chi_{\text{vac}}(\tau, z) + Z_{\text{EG},K3}(\tau, z) \nn \\
&= u^{-1} y^{-1} + u y^{-1} + u^{-1} y + u y + 20 + \big(u^{-1} y^{-1} + u y^{-1} + u^{-1} y + uy \nn\\
&\phantom{aaaa} -130y - 130y^{-1} + 20 y^2 + 20 y^{-2} - 2u - 2u^{-1} + 220\big)q+ \mc{O}(q^2) \nn\\
&:= \sum_{m, \ell, \ell'} c_{K3}(m, \ell, \ell') q^m y^\ell u^{\ell'}
\end{align}
where $Z_{\text{EG},K3}(\tau, z)$ is the elliptic genus of a $K3$ surface and $\chi_{\text{vac}}(\tau,z)$ is the Ramond vacuum character of the $\mc{N}=4$ algebra at $c=6$.\footnote{This is given in (\ref{eq:moose}), by taking $\chi^{s,\text{R}}_0$ with $m=1$.}

\begin{figure}[!h]
\centering
\includegraphics[width=0.85\textwidth]{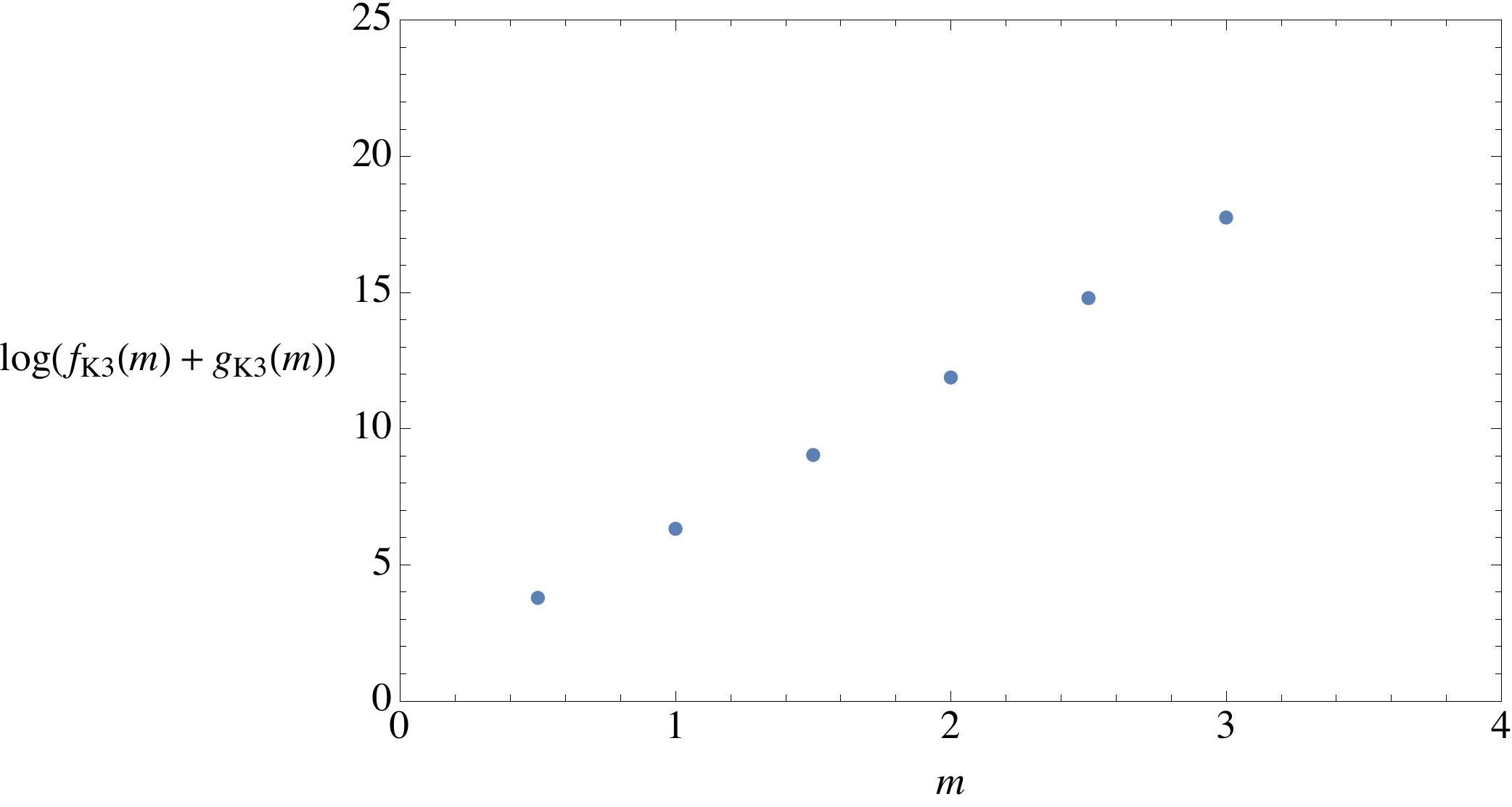}
\caption{A plot of $\log{(f_{K3}(m)+g_{K3}(m))}$ up to $m=3$ for a generic $K3$. Note the exponential growth in $m$.}
\label{fig:squirrel}
\end{figure}

The analysis of the growth follows exactly the same as in Section \ref{sec:t4z2}, with
\be
\frac1NZ_{\text{HEG}, \text{Sym}^N(K3)}^{\text{NS}}(\tau, 1/2, 1/2) = \prod_{m\el \frac{\bb Z}2, m>0} \frac{(1+q^m)^{f_{K3}(m)}}{(1-q^m)^{g_{K3}(m)}} + \mc{O}(q^{N/4}),
\ee
and $f_{K3}(m)$ and $g_{K3}(m)$ defined as in (\ref{eq:ponyup}) (with $c_{X}$ replaced with $c_{K3}$). The first few values are plotted below in Table \ref{tab:rabbit}, and plotted in Figure \ref{fig:squirrel}.
\begin{table}[h]
\begin{center}
\begin{tabular}{| c || c  c  c  c  c  c |}
\hline
$m$ & $\frac12$ & 1 & $\frac32$ & 2 & $\frac52$ & 3 \\ \hline 
$f_{K3}(m)$ & 0 & 264 & 4160 & 71984 & 1328848 & 25602688\\  
$g_{K3}(m)$ & 44 & 292 & 4204 & 72012 & 1328892 & 25602716 \\ \hline 
\end{tabular}
\end{center}
\caption{First few values of $f_{K3}(m)$ and $g_{K3}(m)$, defined in (\ref{eq:ponyup}) for a generic $K3$ surface.}
\label{tab:rabbit}
\end{table}

\subsubsection{Supergravity}

Finally we can also analyze the growth of the Hodge elliptic genus at the supergravity point in moduli space. Here, unfortunately, the conformal field theory becomes very strongly coupled, and intractable; however, we can obtain the BPS spectrum by looking at the supergravity Kaluza-Klein modes. The 6d $\mc{N}=(2,0)$ supergravity KK spectrum on AdS$_3 \times S^3$ was computed in \cite{deBoer:1998kjm, deBoer:1998us}, and is organized into representations of $SU(1,1|2)\times SU(1,1|2)$. The KK spectrum has only short representations of $SU(1,1|2)\times SU(1,1|2)$, which are labelled in \cite{deBoer:1998kjm} by  $(j, j')_S$. This represents a chiral primary state on both the left and right, with $L_0$ eigenvalue $j/2$ and $J_0$ eigenvalue $j$; and $\bar L_0$ eigenvalue $j'/2$ and $\bar J_0$ eigenvalue $j'$. The supergravity multiplet is then obtained by acting with global part of the $\mc{N}=(4,4)$ algebra. In \cite{deBoer:1998us}, agreement with the elliptic genus is obtained by introducing an exclusion principle, which introduces a ``degree" to each multiplet $(j,j')$. At finite $N$, the supergravity quarter-BPS spectrum is given by multiparticle states whose total degree is equal to $N$.

The short multiplets and their degrees are given in Eqn (2.8) of \cite{deBoer:1998us}; we reproduce it below, where each triplet $(j,j';d)_S$ gives the spins of each chiral primary, and the degree:
\begin{align}
&(m-1,m+1;m)_S \nn\\
&(m+1;m+1;m)_S\nn\\
&20(m,m;m)_S\nn\\
&(m,m;m+1)_S\nn\\
&(m+1,m-1;m)_S.
\label{eq:bears}
\end{align}
In (\ref{eq:bears}), $m=1, 2, 3, \ldots$.

To get the full BPS spectrum, we need to include multiparticle states as well. This is given then by
\be
\sum_{n\geq0}Z_{\text{HEG}, \text{sugra,}n}(\tau, z, \nu)p^n = \prod_{n>0,m, \ell, \ell'}\frac{1}{(1-p^nq^my^\ell u^{\ell'})^{c_{\text{sugra}}(n,m,\ell,\ell')}}
\label{eq:mountainlion}
\ee
where $c_{\text{sugra}}(n,m,\ell,\ell')$ is counting single-particle states in the Hilbert space with degree $n$, $L_0$ eigenvalue $m$, $J_0$ eigenvalue $\ell$, and $\bar{J_0}$ eigenvalue $\ell'$, weighted with a $(-1)^F$, and only getting contributions from supersymmetric states on the right.

If we write a generating function for the states contributing from a chiral primary of spin $j$, we get (see Eqn. (2.1) in \cite{deBoer:1998us})
\be
\Tr_{(j)_S}(-1)^F q^{L_0} y^{J_0} = \frac{q^{j/2}}{(1-q)(y-y^{-1})}\((y^{j+1}-y^{j-1})-2q^{\frac12}(y^j-y^{-j})+q(y^{j-1}-y^{1-j})\).
\ee
Now we need to sum over all the left-moving states that appear in (\ref{eq:bears}), and include both the degree and the $u$-dependence from the right-moving ground states. The supergravity calculation is interpreted as the NS-NS sector of the CFT \cite{Maldacena:1998bw}, so to match to the NS-R elliptic genus, we spectral flow the right-movers by 1/2 unit. This is done by shifting the $\bar{J_0}$ eigenvalue by $N$, which we interpret as the degree. This means that for each line in (\ref{eq:bears}), we only get one charge under the $\bar{J_0}$ after spectral flow. Our final expression for $c_{\text{sugra}}(n,m,\ell,\ell')$ is then
\begin{align}
\sum_{n,m,\ell,\ell'} &c_{\text{sugra}}(n,m,\ell,\ell') p^n q^m y^\ell u^{\ell'}  =\frac{1}{(1-q)(y-y^{-1})} \bigg(\frac{(u+u^{-1})p^2}{1-q^{1/2}yp}(y^2 q^{1/2} - 2 yq + q^{3/2}) \nn\\
&-\frac{(u+u^{-1})p^2}{1-q^{1/2}y^{-1}p}(y^{-2}q^{1/2}-2y^{-1}q+q^{3/2})+\frac{(u+u^{-1})p}{1-q^{1/2}yp}(y^3q-2y^2q^{3/2}+yq^2)\nn\\
&-\frac{(u+u^{-1})p}{1-q^{1/2}y^{-1}p}(y^{-3}q-2y^{-2}q^{3/2}+y^{-1}q^2)+\frac{20p}{1-q^{1/2}yp}(y^2q^{1/2}-2yq+q^{3/2})\nn\\
&-\frac{20p}{1-q^{1/2}y^{-1}p}(y^{-2}q^{1/2}-2y^{-1}q+q^{3/2})\bigg)+(u+u^{-1})p.
\label{eq:fish}
\end{align}
The first two terms in (\ref{eq:fish}) correspond to the first and fourth lines of (\ref{eq:bears}); the next two correspond to the second and fifth lines; and the next two correspond to the third line. Note that setting $u=1$ indeed reproduces Eqn (5.9) of \cite{deBoer:1998us}.

To find the unsigned growth of the quarter-BPS states, we can simply take (\ref{eq:mountainlion}), and set $y=u=-1$, as seen from (\ref{eq:horses}). This gives
\begin{align}
\sum_{n\geq0}Z_{\text{HEG,sugra},n}(\tau,1/2,1/2)p^n &= \prod_{n>0,m,\ell,\ell'}\frac{1}{(1-p^nq^m(-1)^{\ell+\ell'})^{c_{\text{sugra}}(n,m,\ell,\ell')}} \nn\\
&= \frac{1}{(1+p)^2}\prod_{n>0,m>0,\ell,\ell'}\frac{1}{(1-p^nq^m(-1)^{\ell+\ell'})^{c_{\text{sugra}}(n,m,\ell,\ell')}}.
\label{eq:sun}
\end{align}
To extract the large $N$ behavior of (\ref{eq:sun}), we can set $p$ to $-1$ in the product. Using the form of $c_{\text{sugra}}$ in (\ref{eq:fish}) we then get at large $N$,
\begin{align}
\frac{1}N Z_{\text{HEG,sugra},N}(\tau, 1/2, 1/2) &= \prod_{n>0,m>0,\ell,\ell'}\frac{1}{(1-q^m(-1)^{n+\ell+\ell'})^{c_{\text{sugra}}(n,m,\ell,\ell')}} \nn\\
&= \prod_{m=1}^{\infty} \frac{(1+q^{m-\frac12})^{48m^2-48m}}{(1-q^{m-\frac12})^{48m^2-48m+44}}\frac{(1+q^{m})^{48m^2-4}}{(1-q^{m})^{48m^2+24}} \nn\\
&= 1 + 44\sqrt{q} + 1106q +20520q^{3/2} + 310735q^2 + \mc{O}(q^{5/2}).
\label{eq:goldfish}
\end{align}

The form of (\ref{eq:goldfish}) is very similar to that of (\ref{eq:fishfish}), but with the analogous $f(m)+g(m)$ growing as $m^2$, so the $q^n$ coefficient grows as $\exp{\(n^{3/4}\)}$.\footnote{In Section \ref{sec:t4z2}, we showed that any symmetric orbifold has $f_X(m) + g_X(m)$ growing exponentially with $m$ which leads to a Hagedorn density of states. Here, near the supergravity point, we find the analogous $f(m)+g(m)$ growing polynomially with $m$, which leads to a sub-Hagedorn density of states. There is no contradiction, of course, because we are not at the orbifold point in moduli space.} In Appendix \ref{app:B}, we show it grows in particular like $\exp{\(\frac{4 \sqrt{2} \pi}{3^{3/4}}n^{3/4}\)}$. Note that this is parametrically faster than the elliptic genus, which grows as $\exp{\(n^{1/2}\)}$, which means that even at the supergravity point, there must be substantial cancellations between the supersymmetric states in the theory. We will explicitly see some of these cancellations in Section \ref{sec:cat}. Finally we note that we can see the $\exp{\(n^{3/4}\)}$ growth in the supergravity KK spectrum from a naive counting argument.\footnote{We thank Christoph Keller for explaining to us this argument.} From dimensional grounds the number of states in the full CFT at the supergravity should grow as $\exp{\(n^{5/6}\)}$ (due to arguments from scaling of the 6d supergravity). However, we count only BPS states, which gives us two constraints: $\bar{h}=\bar{q}/2$ (BPS condition), and $|h-\bar{h}|\leq2$ (absence of higher spin). This brings us down to the $\exp{(n^{3/4})}$ growth seen.

\subsection{$T^4$}

We can repeat the above analysis for the D1/D5 system on $T^4$. Here the elliptic genus vanishes due to right-moving fermion zero-modes (though one can define a modified index that does not vanish \cite{Maldacena:1999bp}).

\subsubsection{Sym$^N(T^4)$}

To get the Hodge elliptic genus at the symmetric orbifold point, we again need the Hodge elliptic genus for the seed theory. In \cite{Kachru:2016igs}, this was computed for a generic $T^4$ as
\begin{align}
Z_{\text{HEG}, T^4}(\tau,z,\nu) &= \(4\frac{\th_1(\tau,z)}{\th_1(\tau)}u_-\)^2 \nn\\
&= u^{-1}y^{-1} + u^{-1}y + uy^{-1} + uy - 2u^{-1} - 2u - 2y^{-1} - 2y + 4 + \nn \\
&\phantom{aaaa} \big(-2u^{-1} y^{-2} - 2u^{-1}y^2 - 2uy^{-2} - 2uy^2 + 8u^{-1}y^{-1} + 8u^{-1}y+8uy^{-1}+8uy \nn\\
&\phantom{aaaa} - 12u^{-1} - 12u - 16y^{-1} - 16y + 4y^{-2} + 4y^2 + 24\big)q + \mc{O}(q^2) \nn\\
&:= \sum_{m,\ell,\ell'} c_{T^4}(m,\ell,\ell') q^m y^\ell u^{\ell'}.
\label{eq:dinosaur}
\end{align}

\begin{figure}[!h]
\centering
\includegraphics[width=0.85\textwidth]{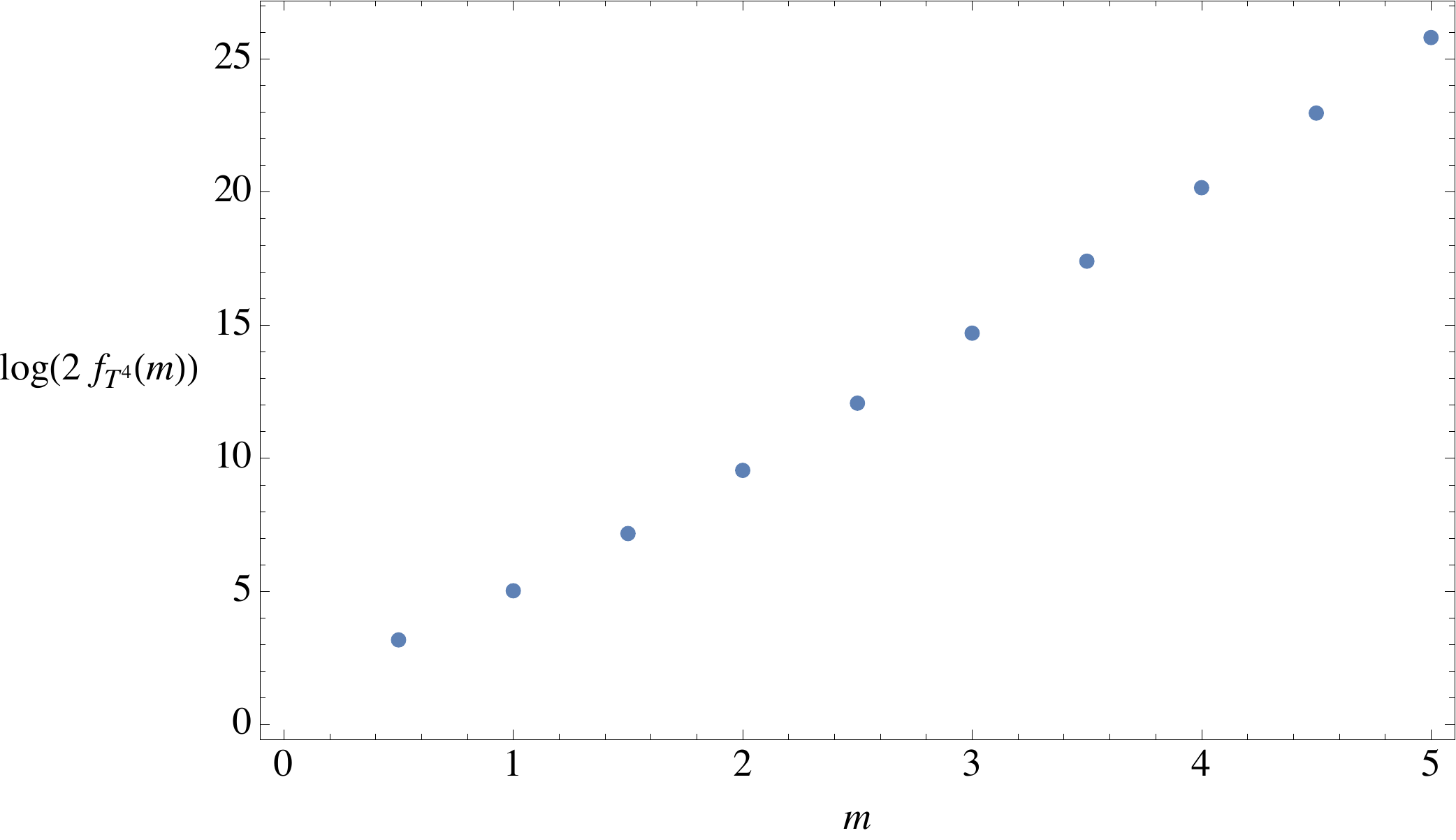}
\caption{A plot of $\log{(2f_{T^4}(m))}$ up to $m=5$ for $T^4$. Note the exponential growth in $m$.}
\label{fig:gg}
\end{figure}

Again, we repeat the analysis in Section \ref{sec:t4z2} to get the growth of states in the Sym$^N(T^4)$ theory. This time however, the RHS of (\ref{eq:us}) vanishes, since all the coefficients $c_{T^4}^{\text{EG}}$ vanish (due to the elliptic genus of $T^4$ being zero). Thus the growth for the quarter BPS spectrum goes as
\begin{align}
\frac1NZ_{\text{HEG}, \text{Sym}^N(T^4)}^{\text{NS}}(\tau, 1/2, 1/2) &= \prod_{\substack{n>0 \\ m\el \frac{\bb Z}2, m>0 \\ m-\frac{\ell}2\el\bb{Z}}} \frac{1}{(1-q^m(-1)^{n+\ell+\ell'} )^{c_{T^4}(nm-\frac{n\ell}2,\ell- n,\ell')}} + \mc{O}(q^{N/4}) \nn \\
&= \prod_{m\el \frac{\bb Z}2, m>0} \(\frac{1+q^m}{1-q^m}\)^{f_{T^4}(m)} + \mc{O}(q^{N/4}).
\label{eq:koala}
\end{align}
The first few values of $f_{T^4}(m)$ are listed in Table \ref{tab:lizard} and plotted in Figure \ref{fig:gg}.
\begin{table}[h]
\begin{center}
\begin{tabular}{| c || c  c  c  c  c  c  c  c |}
\hline
$m$ & $\frac12$ & 1 & $\frac32$ & 2 & $\frac52$ & 3 & $\frac72$ & 4 \\ \hline % & $\frac92$ & 5\\ \hline 
$f_{T^4}(m)$ & 12 & 76 & 652 & 6988 & 87180 & 1207500 & 18021132 & 284382028 \\ \hline% & 4685936140 & 79921293004 \\ \hline 
\end{tabular}
\end{center}
\caption{First few values of $f_{T^4}(m)$ for $T^4$.}
\label{tab:lizard}
\end{table}

\subsubsection{Supergravity}
We can also analyze the 6d $\mc{N}=(2,2)$ supergravity spectrum on AdS$_3 \times S^3$. The KK spectrum will decompose as short representations of $SU(1,1|2)\times SU(1,1|2)$. The table of representations that show, as well as the degree we associate to each, is
\begin{align}
&(m, m; m+1)_S \nn\\
&2(m-1, m;m)_S \nn\\
&2(m,m-1;m)_S \nn\\
&(m-1,m+1;m)_S \nn\\
&(m+1,m-1;m)_S\nn\\
&4(m,m;m)_S\nn\\
&2(m,m+1;m)_S\nn\\
&2(m+1,m;m)_S\nn\\
&(m+1,m+1;m)_S,
\label{eq:caterpillar}
\end{align}
$m=1, 2, \ldots$. (The states with a 2 in front have the highest weight state fermionic, so they will come with a sign when we count.) Counting the single-particle states weighted with a $(-1)^F$ then gives
\begin{align}
\sum_{n,m,\ell,\ell'} &c_{\text{sugra}}(n,m,\ell,\ell') p^n q^m y^\ell u^{\ell'}= \nn\\ &\frac{1}{(1-q)(y-y^{-1})} \bigg(\frac{(u+u^{-1}-2)p^2+(4-2u^{-1}-2u)p}{1-q^{1/2}yp}(y^2 q^{1/2} - 2 yq + q^{3/2}) \nn\\
&-\frac{(u+u^{-1}-2)p^2+(4-2u^{-1}-2u)p}{1-q^{1/2}y^{-1}p}(y^{-2}q^{1/2}-2y^{-1}q+q^{3/2})\nn\\&+\frac{(u+u^{-1}-2)p}{1-q^{1/2}yp}(y^3q-2y^2q^{3/2}+yq^2)-\frac{(u+u^{-1}-2)p}{1-q^{1/2}y^{-1}p}(y^{-3}q-2y^{-2}q^{3/2}+y^{-1}q^2)\bigg)\nn\\&+(u+u^{-1}-2)p.
\label{eq:catfish}
\end{align}

We then get
\begin{align}
\sum_{n\geq0}Z_{\text{HEG,sugra,}n}(\tau,1/2,1/2)p^n &= \prod_{n>0,m,\ell,\ell'}\frac{1}{(1-p^nq^m(-1)^{\ell+\ell'})^{c_{\text{sugra}}(n,m,\ell,\ell')}} \nn\\
&= \frac{(1-p)^2}{(1+p)^2} \prod_{n>0,m>0,\ell,\ell'}\frac{1}{(1-p^nq^m(-1)^{\ell+\ell'})^{c_{\text{sugra}}(n,m,\ell,\ell')}}.
\label{eq:salmon}
\end{align}
Again, we can extract the large $N$ behavior by setting $p=-1$ in the product in (\ref{eq:salmon}). This gives
\begin{align}
\frac1N Z_{\text{HEG,sugra,}N}(\tau,1/2,1/2) &= 4\prod_{n>0,m>0,\ell,\ell'}\frac{1}{(1-q^m(-1)^{n+\ell+\ell'})^{c_{\text{sugra}}(n,m,\ell,\ell')}} \nn\\
&=4\prod_{m=1}^{\infty}\(\frac{1+q^{m/2}}{1-q^{m/2}}\)^{8m^2+4}\nn\\
&= 4 + 96\sqrt{q} + 1440q + 16768q^{3/2}+165024q^2+\mc{O}(q^{5/2}).
\label{eq:yosemite}
\end{align}
As far as asymptotics, the $q^n$ term in (\ref{eq:yosemite}) grows as $\exp{\(\frac{4\pi(2^{3/4})}3 n^{3/4}\)}$. (See Appendix \ref{app:B} for derivation.)\footnote{See also \cite{Mandal:2007ug, Raju:2007uj} for analysis of the quarter-BPS spectrum at the supergravity point.} % X = (2/3)^(1/4) * K3's #

\section{Character Decomposition} 
\label{sec:cat}
In this section we now decompose the various Hodge elliptic genera computed into characters of the $\mc{N}=4$ algebra. These representations and their characters were studied in \cite{Eguchi:1987sm, Eguchi:1987wf, Eguchi:1988af}, we review some results below.

\subsection{$\mc{N}=4$ Characters}
We label the representations of the $\mc{N}=4$ superconformal algebra by the $L_0$ and $J_0$ eigenvalues of their highest weight state ($h$ and $j$ respectively). The representations come in two types: short (or BPS) representations which have $h=\frac{j}{2}$, and long (or non-BPS) representations which have $h>\frac{j}{2}$. There are $m+1$ different short representation, corresponding to $j=0, 1, \ldots, m$, and there are $m$ families of long representations labelled by $j=0, 1, \ldots, m-1$, where $c=6m$. The characters of each representation $r$, defined as
\be
\chi(\tau,z) = \Tr_{r}\((-1)^F y^{J_0} q^{L_0}\),
\ee
are given by
\begin{align}
\chi^{s,\text{NS}}_j(\tau,z) &= q^{j/2} (-1)^{j} \(\frac{-i q^{1/4} \th_4(\tau, z)^2}{\th_1(\tau, 2z)\eta(\tau)^3}\) \nn\\ &\phantom{aaa}\sum_{k\el\bb Z} q^{(m+1)k^2+(j+1)k}\(\frac{y^{2(m+1)k+j+1}}{(1-yq^{k+\frac12})^2} - \frac{y^{-2(m+1)k-j-1}}{(1-y^{-1}q^{k+\frac12})^2}\) \nn \\
\chi^{\ell,\text{NS}}_{j, h}(\tau,z) &= q^{j/2+h}(-1)^{j} \(\frac{-i q^{1/4} \th_4(\tau,z)^2}{\th_1(\tau,2z)\eta(\tau)^3}\) \nn\\& \phantom{aaa} \sum_{k\el\bb Z}q^{(m+1)k^2+(j+1)k}\(y^{2(m+1)k+j+1}-y^{-2(m+1)k-j-1}\)
\label{eq:eagle}
\end{align}
where $\chi^{s,\text{NS}}_j$ is a short representation with highest weight state of spin and weight $j/2$, and $\chi^{\ell}_{j, h}$ is a long representation with highest weight state of spin $j/2$ and weight $j/2+h$. In (\ref{eq:eagle}), all the characters are computed in the NS sector Hilbert space. To get the R sector character, one simply spectral flows by $1/2$ unit. The R sector characters are given by\footnote{For convenience, our convention is that in the R sector, the characters are not defined as $\Tr_{r}\((-1)^F y^{J_0} q^{L_0}\)$, but rather as $\Tr_{r}\((-1)^F y^{J_0} q^{L_0-\frac{c}{24}}\)$. This way, both the NS vacuum character and R vacua characters start at $q^0$. Also note that we label each character by $j$ which is twice the spin in the NS sector. The spin of the highest weight state in the R sector is $(m-j)/2$, not $j/2$.}
\begin{align}
\chi^{s,\text{R}}_j(\tau,z) &= (-1)^{j+m} \frac{i \th_1(\tau, z)^2}{\th_1(\tau, 2z)\eta(\tau)^3} \sum_{k\el\bb Z} \frac{q^{(m+1)k^2+k}y^{2(m+1)k+1}}{(1-yq^k)^2}\(y^{m-j+1} q^{k(m-j+1)} - y^{-(m-j+1)}q^{-k(m-j+1)}\) \nn \\
\chi^{\ell,\text{R}}_{j, h}(\tau,z) &= q^{h}(-1)^{j+m} \frac{i \th_1(\tau,z)^2}{\th_1(\tau,2z)\eta(\tau)^3} \sum_{k\el\bb Z}q^{(m+1)k^2}y^{2(m+1)k}\(q^{k(m-j)}y^{m-j}-q^{-k(m-j)}y^{-(m-j)}\)
\label{eq:moose}
\end{align}

Evaluating the long characters at $z=0$ reduces to calculating the Witten index, which vanishes for long representations, and equals a constant for short representations.
\begin{align}
\chi^{s,\text{R}}_{j}(\tau,0) &= (m+1-j)(-1)^{j+m} \nn \\
\chi^{\ell,\text{R}}_{j,h}(\tau,0) &= 0.
\label{eq:hummus}
\end{align}
Moreover, some short multiplets can combine to form long multiplets. In particular, 
\be
\chi^s_j + 2\chi^s_{j+1} + \chi^s_{j+2} = \chi^{\ell}_{j, 0}
\label{eq:chicken}
\ee
for $j=0, 1, \ldots, m-2$, and
\be
\chi^s_{m-1} + 2\chi^s_m = \chi^{\ell}_{m-1,0}.
\label{eq:turkey}
\ee
Note that (\ref{eq:chicken}), (\ref{eq:turkey}) have vanishing LHS when setting $z=0$, which means the elliptic genus is invariant when these short multiplets pair up.

Since we are decomposing the NS-R Hodge elliptic genus, we will use NS characters on the left, and R characters on the right. Moreover, we only get short representations on the right by definition of the Hodge elliptic genus, which means we only get integral $h$ long representations on the left (by modular invariance). Thus a general decomposition will look like
\be
Z_{\text{HEG}}(\tau,z,\nu) = \sum_{j, \bar{j}} c_{j,\bar{j}} \chi_j^{s,\text{NS}}(\tau,z) \bar{\chi}_{\bar{j}}^{s, \text{R}}(\bar\tau, \nu) + \sum_{j,h,\bar{j}} d_{j,h,\bar{j}} \chi_{j,h}^{\ell,\text{NS}}(\tau,z)\bar{\chi}_{\bar{j}}^{s,\text{R}}(\bar\tau,\nu)\bigg|_{\bar h = \frac c{24}}.
\label{eq:wolf}
\ee

\subsection{$K3$ Decomposition}

Now we will decompose the Sym$^N(K3)$ Hodge elliptic genus into $\mc{N}=4$ characters at the three regions in moduli space we calculated in Section \ref{sec:k3}: Sym$^N(T^4/\bb Z_2)$, Sym$^N(K3)$, and supergravity.

The first observation we make is that the coefficients $c_{j,\bar{j}}$ in (\ref{eq:wolf}) are independent of the moduli. These coefficients count half-BPS states, and are fully determined by the Hodge diamond of Sym$^N(K3)$ which is independent of moduli and given by \cite{Katz:2014uaa}
\be
\sum_{n=0}^{\infty} \chi_{\text{Hodge, Sym}^n(K3)}(z,\nu)p^n = \prod_{k=1}^{\infty} \frac{1}{(1-u^{-1}y^{-1}p^k)(1-u^{-1}yp^k)(1-uy^{-1}p^k)(1-uyp^k)(1-p^k)^{20}}.
\ee

% Say something about decomposition.. maybe (1-q^1/2) * \prod * 1/(1-q^{i/2})^{23}?

The quarter-BPS state character decomposition does depend on moduli. We will look at the characters contributing to low-lying states in the Hodge elliptic genus. In (\ref{eq:wolf}), a long multiplet $\chi_{j,h}^{\ell,\text{NS}}$ starts at $q^{j/2+h}$. The lightest quarter-BPS states thus have $h=1, j=0$. We show the multiplicities of all right-moving short characters multiplying $\chi_{0,1}^{\ell}$ in Table \ref{tab:q12}.

\begin{table}[h]
\begin{center}
\begin{tabular}{| c || c  c  c  c  c  c |}
\hline
Theory & $\chi^{\ell}_{0,1}\bar{\chi}^{s}_0$ & $\chi^{\ell}_{0,1}\bar{\chi}^{s}_1$ & $\chi^{\ell}_{0,1}\bar{\chi}^{s}_2$ & $\chi^{\ell}_{0,1}\bar{\chi}^{s}_3$ & $\chi^{\ell}_{0,1}\bar{\chi}^{s}_4$ & $\chi^{\ell}_{0,1}\bar{\chi}^{s}_{>4}$ \\ \hline
Sym$^N(T^4/\bb Z_2)$ & 3 & 102 & 428 & 142 & 44 & 0 \\ %\hline 
&  & {\footnotesize $\downarrow 3$} & {\footnotesize $\downarrow 6$}  & {\footnotesize $\downarrow 3$}  & & \\ %\hline
Sym$^N(K3)$ & 0 & 90 & 410 & 130 & 41 & 0 \\ %\hline 
&  & & {\footnotesize $\downarrow 90$}  & {\footnotesize $\downarrow 20$}  & & \\ %\hline
Supergravity & 0 & 0 & 210 & 0 & 21 & 0 \\ \hline
\end{tabular}
\end{center}
\caption{Coefficients of all short characters multiplying $\chi^{\ell}_{0,1}$ for Sym$^N(T^4/\bb{Z}_2)$, Sym$^N(K3)$, and the supergravity region. These are the coefficients $d_{0,1,\bar{j}}$ in (\ref{eq:wolf}) for $\bar{j}=0, 1, 2, \ldots N$. A number $n$ next to a downward-pointing arrow below short multiplet $\bar{\chi}_j^{s}$ represents $2n$ short multiplets of type $j$ combining with $n$ of type $j-1$ and $j+1$ to form $n$ long multiplets.}
\label{tab:q12}
\end{table}

We pause to make two points. First, at the symmetric orbifold points, we have many more states than at the supergravity point. This is indeed consistent with what was seen in Section \ref{sec:k3}, where we showed at the symmetric orbifold point, the quarter-BPS states grew exponentially; compared to the subexponential growth at the supergravity region. Second, at the supergravity point, no more cancellations can occur amongst BPS states with $\chi^{\ell}_{0,1}$ on the left. For BPS states to ``pair up", we need two of type $j$ to combine with one of type $j-1$ and $j+1$; this is impossible in the third line of Table \ref{tab:q12}. We will see, however, that this not always true in the supergravity. In Table \ref{tab:sushi}, we show the multiplicities of all right-moving short characters multiplying  $\chi_{0,1}^{\ell}$ ($q$ above the vacuum in the NS sector), $\chi_{1,1}^{\ell}$ ($q^{3/2}$ above the vacuum in the NS sector), $\chi_{2,1}^{\ell}$ ($q^2$ above the vacuum in the NS sector), and $\chi_{0,2}^{\ell}$ ($q^2$ above the vacuum in the NS sector) at the supergravity point. 

\begin{table}[h]
\begin{center}
\begin{tabular}{| c || c  c  c  c  c  c  c  c  c  c  c  c |}
\hline
Long & $\bar{\chi}^{s}_0$ & $\bar{\chi}^{s}_1$ & $\bar{\chi}^{s}_2$ & $\bar{\chi}^{s}_3$ & $\bar{\chi}^{s}_{4}$ & $\bar{\chi}^{s}_{5}$ & $\bar{\chi}^{s}_{6}$ & $\bar{\chi}^{s}_{7}$ & $\bar{\chi}^{s}_{8}$ & $\bar{\chi}^{s}_{9}$ & $\bar{\chi}^{s}_{10}$ & $\bar{\chi}^{s}_{>10}$ \\ \hline
$\chi_{0,1}^{\ell}$ & 0 & 0 & 210 & 0 & 21 & 0 & 0 & 0 & 0 & 0 & 0 & 0 \\ %\hline 
$\chi_{1,1}^{\ell}$ & 0 & 0 & 0 & 3542 & 0 & 484 & 0 & 22 & 0 & 0 & 0 & 0 \\ %\hline 
$\chi_{2,1}^{\ell}$ & 0 & 0 & 21 & 0 & 36961 & 0 & 6281 & 0 & 506 & 0 & 22 & 0 \\ %\hline 
$\chi_{0,2}^{\ell}$ & 0 & 0 & 231 & 2660 & 21526 & 420 & 3796 & 0 & 275 & 0 & 1 & 0\\ \hline
\end{tabular}
\end{center}
\caption{Coefficients of all short characters multiplying the first four long characters that at the supergravity region in Sym$^N(K3)$. Note that cancellations \emph{can} (but do not) occur for $\chi^{\ell}_{0,2}$.}
\label{tab:sushi}
\end{table}

For quarter-BPS states with $\chi^{\ell}_{0,2}$ on the left, we see from Table \ref{tab:sushi} that cancellations can occur but do not (for instance, $\bar\chi_2^s$, 2$\bar\chi_3^s$, and $\bar\chi_4^s$ can pair up 231 times). In fact, this appears to be a general statement for all states with $\chi_{j,h>1}^{\ell}$ on the left.\footnote{In fact, we knew that at the supergravity point, we had to have cancellations possible in the character decomposition -- if cancellations could never occur, then the growth of the total number of quarter BPS-states would be the same as the growth of the signed sum of quarter BPS-states. But in fact the former grows as $\exp{\(n^{3/4}\)}$ and the latter as $\exp{\(n^{1/2}\)}$.} Assuming that supergravity has the slowest growth of low-lying states in the moduli space of Sym$^N(K3)$, this implies that at a generic point in the moduli space, it is \emph{not} the case that short multiplets that can pair up always do pair up. It would be interesting to understand if there was, e.g. an extra symmetry that protected some of the short multiplets that do not pair up in Table \ref{tab:sushi}. Another logical possibility is that there is a point in the moduli space with slower growth than supergravity, in which all short multiplets that can cancel do cancel.

Finally we end this section with a curious numerological observation. Many of the multiplicities in Table \ref{tab:sushi} decompose very nicely under sizes of irreducible representations of the sporadic Mathieu group $M_{22}$. For instance, 
\begin{align}
210 &= {\bf 210},~ 21 = {\bf 21},~ 484 = {\bf 99} + {\bf 385},~ 22 = {\bf 1} + {\bf 21},~231={\bf 231}\nn\\506 &= {\bf 1} + {\bf 21} + {\bf 99} + {\bf 385},~420 = {\bf 210} + {\bf 210},~275 = {\bf 1} + {\bf 1} + {\bf 21} + {\bf 21} + {\bf 231}.
\label{eq:crackpot}
\end{align}
where the bolded numbers in (\ref{eq:crackpot}) are irreducible representations of $M_{22}$. It would be interesting if there were some CFT in the moduli space with a natural $M_{22}$ symmetry (for instance, some point in the moduli space where the supergravity interactions break a naive symmetry of the states in the 6d $(2,0)$ supergravity theory into an $M_{22}$). Relations between the elliptic genus of $K3$ and the related Mathieu group $M_{24}$ have been discussed in many papers, starting with \cite{Eguchi:2010ej}.

\section{Discussion}
\label{sec:fish}

In this paper, we analyzed the recently introduced Hodge elliptic genus \cite{Kachru:2016igs} at various points in the moduli space of a 2d CFT with target Sym$^N(K3)$ and Sym$^N(T^4)$. We showed that at the symmetric orbifold point of any supersymmetric sigma model with target Sym$^N(X)$, the entropy of the low-lying quarter-BPS spectrum grows exponentially with the dimension. However, after a deformation to the large-radius supergravity point for both $K3$ and $T^4$, the total entropy of the quarter-BPS spectrum scales as $\exp{\(n^{3/4}\)}$. This is to be contrasted with the signed quarter-BPS spectrum (the elliptic genus), whose growth scales as $\exp{\(n^{1/2}\)}$. This means that at the supergravity point, there are still many cancellations that do occur between BPS states.

We can make this more precise by looking at the character decomposition of this quantity. We decompose the Hodge elliptic genera computed at various points in the Sym$^N(K3)$ moduli space into $\mc{N}=4$ characters. As we move from the symmetric orbifold point to the large radius supergravity, we can explicitly see short multiplets pair up to form long multiplets, which vanish in the genus (see Table \ref{tab:q12}). However, even at the supergravity point, we see many short multiplets that could potentially pair up that do not. Finally, we note that at the supergravity point, the Hodge elliptic genus decomposed into $\mc{N}=4$ characters suggest a possible relation to the sporadic group $M_{22}$. We conclude with a list of potentially interesting questions:
\begin{itemize}
\item Is the growth exhibited at the supergravity point generic in the Sym$^N(K3)$ and Sym$^N(T^4)$ moduli space? If not, is there a point that grows slower than supergravity?
\item Is there some extra symmetry preventing more quarter-BPS states from combining into nonsupersymmetric states at a generic point? 
\item Is there any relation between the group $M_{22}$ and the $D1/D5$ system?
\item Do we get anything interesting studying the decomposition of Sym$^N(T^4)$ Hodge elliptic genera into contracted large $\mc{N}=4$ characters?
\item Is there an interpretation for places in the moduli space where the Hodge elliptic genus ``jumps"?
\item Can the Hodge elliptic genus provide a more refined count to the black hole entropy when the elliptic genus gets the count wrong due to too many cancellations (see e.g. \cite{Vafa:1997gr})?
\item Can cancellations at generic points in the Hodge elliptic genus be used to understand the BPS spectrum in the $S$-dual of the D1/D5 system \cite{Seiberg:1999xz}?\footnote{See also \cite{Mandal:2007ug, Raju:2007uj}.} 
%\item Who \emph{really} killed JFK? And why are there so many holes in the official moon landing story?
\end{itemize}

{\centerline{\bf{Acknowledgements}}

\medskip
It is a pleasure to thank Ethan Dyer, Shamit Kachru, Christoph Keller, Suvrat Raju, Shu-Heng Shao, Roberto Volpato, and Kenny Wong for very useful discussions, as well as Shamit Kachru and Christoph Keller for very helpful comments on a draft. We thank Katrin Wendland for kindly allowing use of her unpublished note. This work was supported by a Stanford Graduate Fellowship and an NSF Graduate Fellowship.

\appendix

\section{Theta Functions}
\label{app:A}

We define the standard Jacobi theta functions
\begin{align}
\th_1(\tau, z) &= -i q^{\frac18} y^\frac12 \prod_{n=1}^{\infty} (1-q^n)(1-yq^n)(1-y^{-1}q^{n-1}) \nn \\
\th_2(\tau, z) &= q^\frac18 y^\frac12 \prod_{n=1}^{\infty} (1-q^n)(1+yq^n)(1+y^{-1}q^{n-1}) \nn \\
\th_3(\tau, z) &= \prod_{n=1}^{\infty} (1-q^n)(1+yq^{n-\frac12})(1+y^{-1}q^{n-\frac12}) \nn \\
\th_4(\tau, z) &= \prod_{n=1}^{\infty} (1-q^n)(1-yq^{n-\frac12})(1-y^{-1}q^{n-\frac12}).
\end{align}

With the second argument suppressed, we are taking it at $z=0$, except for $\th_1$, where we remove the zero-mode. More explicitly
\begin{align}
\th_1(\tau) &= -2i q^{\frac18} \prod_{n=1}^{\infty} (1-q^n)^3 \nn \\
\th_2(\tau) &= \th_2(\tau, 0) = 2q^\frac18 \prod_{n=1}^{\infty} (1-q^n)(1+q^n)^2 \nn \\
\th_3(\tau) &= \th_3(\tau, 0) = \prod_{n=1}^{\infty} (1-q^n)(1+q^{n-\frac12})^2 \nn \\
\th_4(\tau) &= \th_4(\tau, 0) = \prod_{n=1}^{\infty} (1-q^n)(1-q^{n-\frac12})^2.
\end{align}

Finally we define the Dedekind eta function as
\be
\eta(\tau) = q^{1/24} \prod_{n=1}^{\infty}(1-q^n).
\ee

\section{Derivation of prefactor}
\label{app:B}

In this appendix we derive the growth of the $q^n$ coefficient of
\be
\prod_{m=1}^{\infty} \frac{1}{(1-q^m)^{a m^p}}
\label{eq:bosons}
\ee
and
\be
\prod_{m=1}^{\infty} (1+q^m)^{a m^p}
\label{eq:fermions}
\ee
which has been used often in Section \ref{sec:dog}. Let's first consider (\ref{eq:bosons}) where we take $q=1-\epsilon$:
\begin{align}
\prod_{m=1}^{\infty} \frac1{(1-q^m)^{a m^p}} &= \exp{\(-\sum_{m=1}^{\infty} a m^p \log{(1-q^m)}\)} \nn \\
&= \exp{\(\sum_{m=1}^{\infty}\sum_{n=1}^{\infty} \frac{a m^p}n q^{nm} \)} \nn \\
&\sim \exp{\(\sum_{n=1}^{\infty} \frac{a p!}{n(1-q^n)^{p+1}}\)}\nn\\
&\sim \exp{\(\sum_{n=1}^{\infty} \frac{a p!}{n(n \epsilon)^{p+1}}\)}\nn\\
&\sim\exp{\(\frac{a p! \zeta(p+2)}{(-\log{q})^{p+1}}\)}.
\end{align}
The $q^x$ coefficient is given by doing the integral
\begin{align}
\frac{1}{2\pi i} \oint dq \exp{\(\frac{a p! \zeta(p+2)}{(-\log{q})^{p+1}} - x\log q\)}
\end{align}
which can be evaluated by saddle to give
\be
a^{\frac1{p+2}} (p!)^{\frac1{p+2}} \zeta(p+2)^{\frac1{p+2}}\(\frac{p+2}{(p+1)^{\frac{p+1}{p+2}}}\) x^{\frac{p+1}{p+2}}.
\label{eq:gross}
\ee
We can evaluate the $q^x$ growth of (\ref{eq:fermions}) using the same strategy; the final answer gives
\be
(1-2^{-p-1})^{\frac1{p+2}}a^{\frac1{p+2}} (p!)^{\frac1{p+2}} \zeta(p+2)^{\frac1{p+2}}\(\frac{p+2}{(p+1)^{\frac{p+1}{p+2}}}\) x^{\frac{p+1}{p+2}}.
\label{eq:gross2}
\ee
This is the same as (\ref{eq:gross}), but with $a\rightarrow (1-2^{-p-1})a$. In (\ref{eq:goldfish}), we can therefore take (\ref{eq:gross}) with $a=96+96(1-2^{-p-1})$ and $p=2$, giving $\frac{4\sqrt 2 \pi}{3^{3/4}}$ for the prefactor in the exponential. Similarly, in (\ref{eq:yosemite}), we can take (\ref{eq:gross}) with $a=64+64(1-2^{-p-1})$ and $p=2$, giving $\frac{4\pi(2^{3/4})}3$ as the prefactor in the exponential.

\bibliographystyle{JHEP}
\bibliography{refs}
\end{document}